\documentstyle[aps,prd,preprint,epsf,floats]{revtex}

\preprint{\vbox{\hbox{BIHEP-EP1-98-06\hfill}
                \hbox{UH511-923-99\hfill}}}

\tighten
\newcommand{\etal}{\it et al. \rm}
\newcommand{\rt}{\rightarrow}
\newcommand{\pipi}{\pi^+ \pi^-}
\title{ Charmonium Decays to Axialvector plus Pseudoscalar Mesons}

\author{
J.~Z.~Bai,$^1$   Y.~Ban,$^5$      J.~G.~Bian,$^1$
I.~Blum,$^{12}$ 
G.~P.~Chen,$^1$  H.~F.~Chen,$^{11}$  
J.~Chen,$^3$ 
J.~C.~Chen,$^1$  Y.~Chen,$^1$ Y.~B.~Chen,$^1$
% ...
\\ (BES Collaboration)}

\address{
$^1$Institute of High Energy Physics, Beijing 100039, People's Republic of
 China\\
$^2$California Institute of Technology, Pasadena, California 91125\\
$^3$Colorado State University, Fort Collins, Colorado 80523\\
$^4$Hangzhou University, Hangzhou 310028, People's Republic of China\\
$^5$Peking University, Beijing 100871, People's Republic of China\\
$^6$Shandong University, Jinan 250100, People's Republic of China\\
$^7$Shanghai Jiaotong University, Shanghai 200030, People's Republic of China\\
$^8$Stanford Linear Accelerator Center, Stanford, California 94309\\
$^9$University of Hawaii, Honolulu, Hawaii 96822\\
$^{10}$University of California at Irvine, Irvine, California 92717\\
$^{11}$University of Science and Technology of China, Hefei 230026,
People's Republic of China\\
$^{12}$University of Texas at Dallas, Richardson, Texas 75083-0688}

\date{\today}
\begin{document}
\maketitle

\begin{abstract}

A sample of 3.79 million $\psi(2S)$ events
is used to study the decays of charmonium to
axialvector plus pseudoscalar
mesons.  The branching fraction 
for the decay $\psi(2S)\rt b^{\pm}_1(1235) \pi^{\mp}$
agrees with expectations based on scaling 
the corresponding $J/\psi$ branching fraction.
Flavor-SU(3)-violating $K_1(1270)$-$K_1(1400)$ asymmetries 
with opposite character for $\psi(2S)$ and $J/\psi$ decays
are observed. This contrasting behavior 
can not be accommodated by adjustments of 
the singlet-triplet mixing angle.

\end{abstract}

\vspace{0.25in}



In perturbative QCD, the dominant process for
hadronic decays of both the $J/\psi$ and 
the $\psi(2S)$ is 
annihilation into three gluons followed by the
hadronization of these gluons into
physically observable hadrons.  The similarity
between the parton-level final states has led to
the conjecture that the ratio of the $J/\psi$ and $\psi(2S)$ decay
branching fractions into any exclusive final state $X_h$
is given by the ratio of the square of the wave function at the origin
of the constituent $c\overline{c}$ quark state, which is well
determined from the dilepton decay rates~\cite{aaa}:
\[
\frac{{\cal B}[\psi(2S)\rt X_h]}{{\cal B}[J/\psi\rt X_h]} \simeq 
\frac{{\cal B}[\psi(2S)\rt e^+ e^-]}{{\cal B}[J/\psi\rt e^+ e^-]} = 
0.141 \pm 0.012.
\]
This conjecture is sometimes referred to as {\em the 14\% rule}~\cite{12per}.
Although this conjecture seems to work reasonably
well for a number of decay channels,  it fails badly in
the case of  $\psi(2S)$ two-body decays to 
vector plus pseudoscalar meson ($VP$) final states---the
decay $\psi(2S)\rightarrow\rho\pi$ is suppressed relative
to the 14\% rule expectation by more than a 
factor of fifty~\cite{MARKII,BES1}.  This conundrum is
commonly called the $\rho\pi$ puzzle~\cite{BLT}. 
In addition, the
BES group has reported suppressions by factors of at least three in the
vector plus tensor meson ($VT$) final states: $K^*\overline{K_2^*},$
$\rho a_2$, $\omega f_2$ and $\phi f^{\prime}_2$~\cite{BES2}.  To date, 
no convincing evidence
has been uncovered for hadronic $\psi(2S)$ decays that are
enhanced relative to the 14\%-rule expectation.
Since at least one  explanation for the $\rho \pi$ puzzle 
involves a mechanism that suppresses all $\psi(2S)$ 
decays to lowest lying two-body mesons final states~\cite{CT},
it is useful to examine all possibilities.
Here we report first measurements of $\psi(2S)$ decays to
axialvector plus pseudoscalar mesons. 


There are two lowest-lying axialvector-meson octets. These correspond to
the singlet ($^1P_1$) and triplet ($^3P_1$) spin configurations of two 
quarks in a P-wave orbital angular momentum state.
The non-strange, isospin $I=1$ members of the two octets have 
opposite $G$-parity: the $b_1(1235)$ is in the
$^1P_1$ octet and has $G=+1$, while the $a_1(1260)$ is in the
$^3P_1$ octet and has $G=-1$.  Since strong decays of the
$J/\psi$ and $\psi(2S)$ conserve $G$-parity, decays
to the axialvector pseudoscalar
($AP$) pair $b_1\pi$ are allowed and 
seen in $J/\psi$ decays; decays to $a_1\pi$ final states are
forbidden and not seen in $J/\psi$ decays.

The  strange members of the $^3P_1$ and $^1P_1$
octets, the 
$K_A$ and  $K_B$, respectively, are mixtures of
the observed physical states, the $K_1(1270)$ and the
$K_1(1400)$, where
\begin{eqnarray}
  K_A &=& \cos\theta K_1(1400) + \sin\theta K_1(1270)\\
     K_B &=& \cos\theta K_1(1270) - \sin\theta K_1(1400),
\end{eqnarray}
and the mixing angle is near $\theta\simeq 45^0$~\cite{MIX}.
The dominant $K_1(1270)$ decay mode is to $K\rho$ 
(Br = $42 \pm 6 \%$);
the $K_1(1400)$ decays almost always to $K^*\pi$ (Br = $ 94 \pm 6 \%$).

In the limit of strict flavor-SU(3) symmetry,  the amplitudes 
for two-body decays to conjugate mesons in the same 
pair of octets should be equal.  Thus,
since decays to $a_1\pi$ are forbidden by $G$-parity,
decays to $K_A\overline{K}$  are disallowed
by SU(3) and one expects relatively pure $K_B\overline{K}$ 
final states in $J/\psi$ and $\psi(2S)$ decays.  And,
since $\theta\simeq 45^0$, there should be
roughly equal amounts of $K_1(1270)$ and $K_1(1400).$

This analysis is based on a sample of $(3.79 \pm 0.31)$ million 
$e^+e^- \rt \psi(2S)$ events~\cite{BES4},
collected in the BES detector at the BEPC storage ring. 
The BES detector is described in some detail in 
ref.~\cite{BES3}.
The features that are most important for the analysis reported here
are the 40-layer main cylindrical drift chamber (MDC), the 48-scintillation
counter time-of-flight (TOF) system, and
the 12-layer lead-gas barrel electromagnetic shower counter (BSC).  These
are all situated in a 0.4 T solenoidal magnetic field.  Charged particle
track trajectories are measured in the MDC with a momentum resolution of 
$\sigma_p/p = 1.7\% \sqrt{1 + p^2}$ ($p$ in GeV).
The directions and energies of high energy $\gamma$-rays 
are measured in the BSC
with angular and energy
resolutions of $\sigma_{\phi} = $ 4.5 ~mrad, 
$\sigma_{\theta} = $ 12~mrad and
$\sigma_E/E = 0.22/ \sqrt{E}$ ($E$ in GeV), respectively.
We restrict our analysis to photons and
charged tracks that
are in the polar angle region $|cos\theta|<0.80$.
For hadron tracks the time resolution of the barrel TOF is about 450 ps
and the $dE/dx$ resolution is 
about $11\%$, allowing for a $\pi/K $ separation up to 600 MeV. 
For each charged track that passed the kinematic fit, 
the $dE/dx$ and TOF information is
used to determine the probability that the track is 
consistent with being a kaon;
tracks with a probability of more than 10\% are
considered candidate kaons, otherwise they are considered
to be pions. 



\begin{figure}[htp]
\centerline{\epsfysize 3.2 truein
\epsfbox{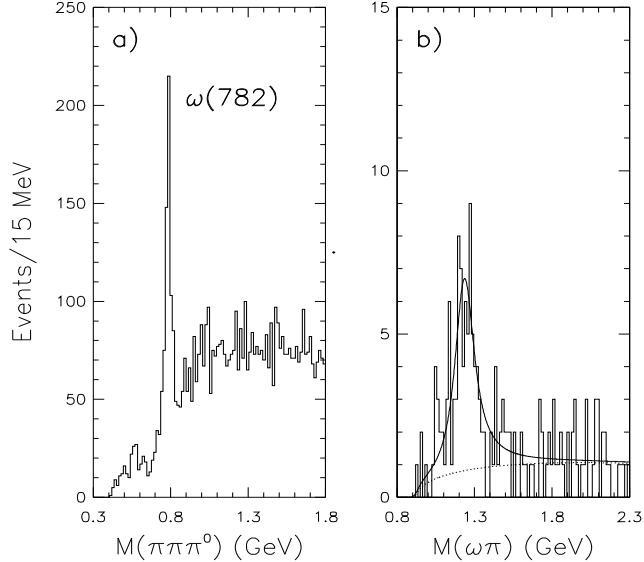}}
\caption{ \label{omega}
The {\bf a)} $\pi^+\pi^- \pi^0$
and {\bf b)} $\omega\pi^{\pm}$ mass distributions
from $\psi(2S)\rightarrow\pi^+\pi^-\pi^+\pi^-\pi^0$ events.
The curve in {\bf b)} is a fit to the $b_1$ resonance
plus a smooth background function.} 
\end{figure}

Since the dominant decay mode of the $b_1$ is $b_1\rightarrow\omega\pi$,
we apply a five constraint kinematic fit to events of the type
$ \psi(2S) \rt \pi^+ \pi^-\pi^+\pi^- \gamma \gamma $,
where the $\gamma\gamma$ invariant mass is
further constrained to be equal to $M_{\pi^{0}}$.  
The $\pi^+\pi^-\pi^0$ mass distribution for events that 
pass the 5-C fit is shown in Fig.~\ref{omega}a,
where there is a prominent peak.
The peak is well fit with a Breit Wigner shape with mass 
and width of the $\omega(782)$ convoluted with a gaussian 
resolution function with $\sigma=9.6$ MeV.
We  identify the best $\pi^+\pi^-\pi^0$ 
combination with invariant mass in the range
$ M_{\omega}\pm 30$~MeV as an
$\omega$ candidate.  
Figure~\ref{omega}b shows the
$\omega \pi^\pm$ mass distribution for events where the
recoiling $\pi^+\pi^-$ pair has an invariant mass
greater than 1.55~GeV.  The latter requirement reduces the 
contamination from $\omega f_2$ final states.
The peak in Fig.~\ref{omega}b is well fit with an S-wave Breit Wigner
function with mass and width fixed at the PDG values for the
$b_1$  ($M_{b_{1}} = 1.232$ and $\Gamma_{b_{1}} = 0.142$~GeV) and 
a background shape
that has a phase-space behavior at threshold that
evolves to a constant level at higher masses.
There are $79.8\pm 12.1$ events in the fitted $b_1$ meson signal 
peak~\cite{resol}.

Using the detection efficiency of $ 0.046 \pm 0.003 $, 
which was determined from a Monte Carlo simulation,
we measure a branching fraction of~\cite{b1}:
\[
{\cal B}(\psi(2S) \rt b^{\pm}_1 \pi^{\mp}) = 
 (5.2 \pm 0.8 \pm 1.0) \times 10^{-4},
\]
where the first error is statistical and the second is
systematic~\cite{system}. 
The result is higher than, but consistent with, the 
14\% rule expectation
applied to the PDG result for the $J/\psi$~\cite{SLO}.


For the $K_1 \overline{K}$ decays, we 
select events of the type $\psi(2S) \rt K^+ K^- \pi^+ \pi^-$ on
the basis of the quality of a 4-C kinematic fit.
This final state includes the dominant $K^{\pm}_1(1270)$ and  
$K^{\pm}_1(1400)$ decay
channels.  
We identify $ \pi^{+} \pi^{-} $ pairs with invariant mass
in the range $M_{\rho} \pm 150$~MeV
as $\rho(770)$ candidates and $K^{\pm}\pi^{\mp}$ pairs with
invariant mass in the range $M_{K^*} \pm 50$~MeV as 
$K^*(892)$ candidates.

\begin{figure}[htp]
\centerline{\epsfysize 3.2 truein
\epsfbox{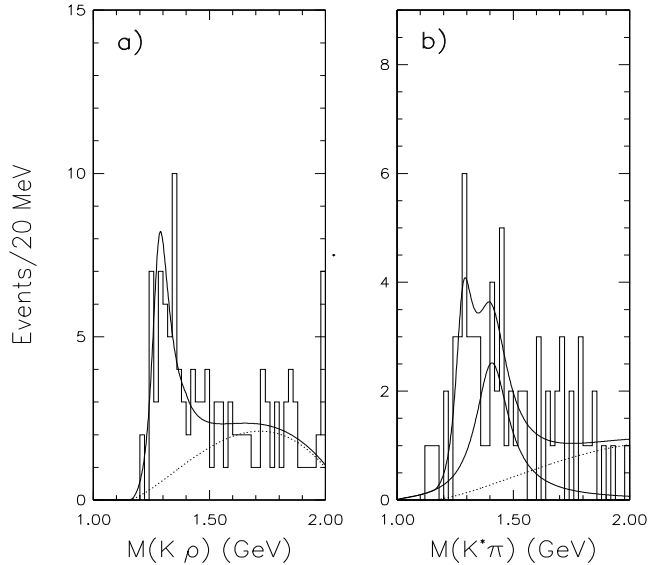}}
\caption{ \label{p_k1k_fig2}
The {\bf a)} $K^{\pm} \rho^0$ and {\bf b)}
$K^{*0}\pi^{\pm}$ mass distributions from
$\psi(2S)\rightarrow K^+K^-\pi^+\pi^-$ events.
Note the difference in the vertical scales.
The curves are the results of the fits
 discussed in the text.}
\end{figure}

The $K^{\pm}\rho$ mass distribution exhibits a strong enhancement
near $M_{K\rho}=1.27$~GeV, as shown in Fig.~\ref{p_k1k_fig2}a.
We fit the $K^\pm \rho^0$ mass distribution  
with a specially devised function, $f_{K \rho},$ 
that takes into account the distortions to the line shape
caused by the restricted phase space 
available for the $K_1(1270) \rt K \rho$ decay~\cite{WMD}.
This plus a smooth background function
that has a phase-space behavior near threshold
provide an adequate fit to the data for masses below 2.0 GeV
and yields a $K_1(1270)$ signal of $53.5\pm 9.5$ events~\cite{resol}.
Using the detection efficiency of 
$ 0.085 \pm 0.012 $,
we determine the branching fraction result of~\cite{contamination}:
\[
{\cal B}(\psi(2S) \rt K^{\pm}_1(1270) K^{\mp}) 
= (10.0 \pm 1.8 \pm 2.1) \times 10^{-4}.
\]

In the $K^{*}\pi^{\pm}$ invariant mass distribution,
shown in Fig.~\ref{p_k1k_fig2}b,
there is no evidence for
a $K_1(1400)$ signal.
Since the $K\rho$ and $K^*\pi$ selection cuts are
not mutually exclusive, some feedthrough from 
$K_1(1270)\rightarrow K\rho$
into the $K^*\pi$ channel is expected, and seen.
The smooth curve in Fig.~\ref{p_k1k_fig2}b
is the result of a fit 
using $f_{K \rho}$ for the $K_1(1270)$,  an
S-wave Breit Wigner with mass and width fixed at the PDG values for the 
$K_1(1400)$ and a smooth
background shape as was used for the $ K \rho$
distribution.
Since the $K^*\pi$ mass distribution 
can be adequately fit without any $K_1(1400)$, 
the resulting $29.8 \pm 9.2$  
$K_1(1400)\rightarrow K^*\pi$ events
and the efficiency of $ 0.090 \pm 0.012 $ 
are used to derive a 90\% CL upper limit of~\cite{limit}:
\noindent
\[
{\cal B}(\psi(2S) \rt K^{\pm}_1 (1400) K^{\mp}) 
< 3.1 \times 10^{-4}~~~90\%~{\rm CL}.
\]
\noindent
Contrary to flavor-SU(3) expectations, 
the $\psi(2S)\rightarrow K_1(1400)\overline{K}$ branching fraction
is smaller than that for the
$K_1(1270)\overline{K}$ channel by at least a factor of three.
To accommodate this with the mixing
angle,  a value $\theta < 29^0$ would be required.


In the absence of any published results for $J/\psi$ decays to these
channels, 
we used the $\psi(2S)\rt\pipi J/\psi$ cascade
events in our $\psi(2S)$ data sample to make a first measurement of
the branching fractions for $J/\psi \rt K_1(1270)\overline{K}$ 
and $K_1(1400)\overline{K}$.
We select events that fit a five-constraint fit to the 
$\psi(2S) \rt \pipi J/\psi; J/\psi\rt K^+ K^- \pi^+ \pi^-$
hypothesis.  We use the particle species 
assignment that gives the best $\chi^2$ value, 
and we use the same $K\rho$ and $K^*\pi$
event selection criteria that are used for the
analysis of direct $\psi(2S)$ decays.

\begin{figure}[htp]
\centerline{\epsfysize 3.2 truein
\epsfbox{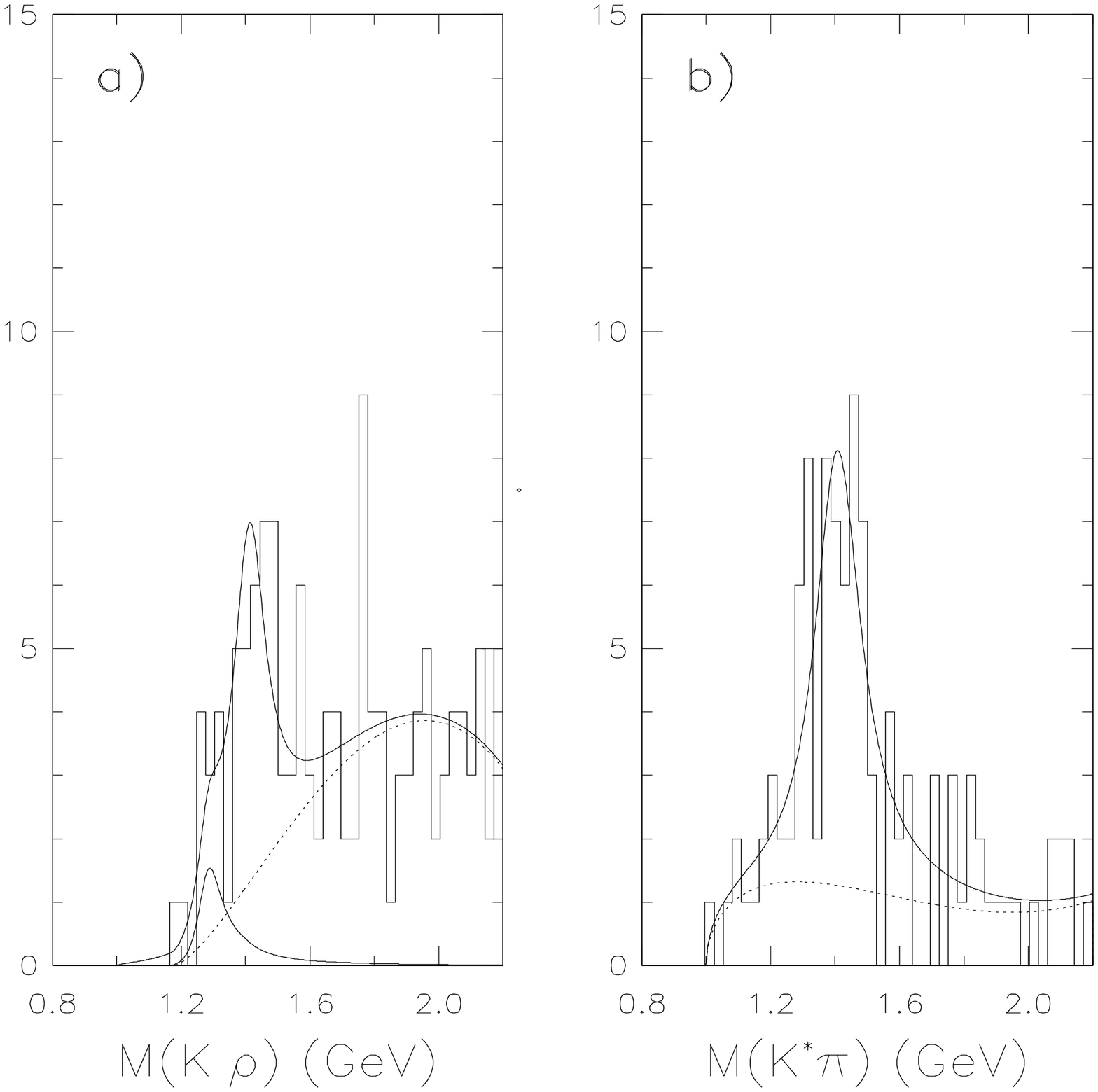}}
\caption{ \label{j_k1k_fig4}
The {\bf a)} $K^{\pm} \rho^0$ and {\bf b)} $K^{*0}\pi^{\pm}$ 
mass distributions from $J/\psi\rightarrow K^+K^-\pi^+\pi^-$ decays.
The curve is the result of the fit described in the text. 
A $K_1(1270)$ signal is not needed to get an acceptable fit to the
$K^{*0}\pi^{\pm}$ distribution.}
\end{figure}

In contrast to the case for the $\psi(2S)$,
the $K\rho$ mass spectrum in $J/\psi\rightarrow K^+K^-\pi^+\pi^-$ decays, 
shown in Fig.~\ref{j_k1k_fig4}a, 
has little
evidence for the $K_1(1270).$  

\vspace{-0.15in}

The small $K_1(1270)$ signal of
$7.7 \pm 5.8~K_1(1270)$ events~\cite{1400fdt} and the efficiency of 
$ 0.025 \pm 0.004$
are used to infer
a 90\% CL upper limit of~\cite{contamination,limit}:
\[
{\cal B}(J/\psi\rt K^{\pm}_1(1270)K^{\mp}) < 3.0 \times 10^{-3}~~90\%~{\rm CL};
\]
this is more than a factor of two below the result expected from applying
the  14\% rule  to our result for $\psi(2S)$ decays to this channel. 

In further contrast to the $\psi(2S)$, 
the $K^{*0} \pi^{\pm}$ mass distribution for the $J/\psi$ decays,
shown in Fig.~\ref{j_k1k_fig4}b, exhibits a clear $K_1(1400)$ signal;
the fit to the  $K^*\pi$ mass spectrum yields $59.0\pm 13.1$ events 
in the $K_1(1400)$ signal~\cite{resol}. The related efficiency is 
$0.030 \pm 0.004$. We find:
\[
{\cal B}(J/\psi\rt K^{\pm}_1(1400)K^{\mp}) 
= (3.8 \pm 0.8 \pm 1.2) \times 10^{-3},
\]
which is above our upper limit for the $K_1(1270)\overline{K}$ mode,
indicating a flavor-SU(3) violation in $J/\psi$ decays that
is opposite to that seen in $\psi(2S)$ decays.  Accommodating
this effect in $J/\psi$ decays by adjusting the mixing angle would
require a value of $\theta > 48^0$, in contradiction to the
$\theta <29^0$ result from $\psi(2S)$ decays.


In conclusion, we report first measurements for the 
$\psi(2S)\rt b^{\pm}_1\pi^{\mp}$ and $K^{\pm}_1(1270)K^{\mp}$ 
decay branching fractions and a 90\% CL upper limit for
${\cal B}(\psi(2S)\rt K_{1}^{\pm}(1400)K^{\mp}).$ 
We find that two of the $AP$ decays are relatively
strong exclusive hadron channels for the 
$\psi(2S).$   
In addition, we report the first observation of the
$J/\psi\rt K^{\pm}_1(1400)K^{\mp}$ decay mode and 
a 90\% CL upper limit for
$J/\psi\rt K^{\pm}_1(1270)K^{\mp}$.

The $\psi(2S)\rightarrow
K_1(1270)\overline{K}$ result is the first observation of an
exclusive $\psi(2S)$ two-body meson 
decay process that is enhanced relative to
the $J/\psi$ in the context of the 14\% rule.  
This result as well as the lack of suppression
in the $b_1\pi$ channel rule out explanations for the $\rho \pi$
puzzle that suppress all $\psi(2S)$ decays to lowest
lying two-body meson final states.
In addition, we observe 
flavor-SU(3)-violating $K_1(1270)$-$K_1(1400)$ asymmetries that have
opposite character for the $\psi(2S)$ and $J/\psi.$  This can not be
accommodated by adjustments of the singlet-triplet mixing angle~\cite{Mahiko}.


We acknowledge the strong efforts of the BEPC staff and the 
helpful assistance we received from the members of the IHEP 
computing center.  One of us (D.P.) thanks M. Suzuki, 
and F. Liu for helpful discussions.
The work of the BES Collaboration is supported in part by
the National Natural Science Foundation of China
under Contract No. 19290400 and the Chinese Academy of Sciences
under contract No. H-10 and E-01 (IHEP), 
and by the Department of
Energy under Contract Nos. DE-FG03-92ER40701 (Caltech),
DE-FG03-93ER40788 (Colorado State University), DE-AC03-76SF00515 (SLAC),
DE-FG03-91ER40679 (UC Irvine), DE-FG03-94ER40833 (U Hawaii),
DE-FG03-95ER40925 (UT Dallas).




\begin{references}
\bibitem{12per} If the change in the QCD coupling strength $\alpha_s(Q^2)$ 
from $Q^2=M^2_{J/\psi}$ to $M^2_{\psi(2S)}$ is considered, the expectation 
for this ratio is lowered to $ 0.12 \pm 0.02$.
\bibitem{resol} The experimental mass resolution  ($\sigma$)
for $b_{1} \rt \omega\pi$ is 12  MeV; for 
$K_{1}(1270)\rt K\rho$ is 17 MeV; and 
for $K_{1}(1400)\rt K^*\pi$ is 18 MeV.
\bibitem{b1} We assume that $b^+_1\rightarrow \omega\pi^+$
branching fraction is 100\%.
\bibitem{system} The systematic errors for all
branching fractions in this paper include the error
on the number of $\psi(2S)$ and the errors on the 
PDG branching fractions for all the intermediate 
channels to the final states. Uncertainties associated with 
the continuum contribution, the Monte Carlo simulation, 
particle identification, and the choice of cuts
are also taken into account.
\bibitem{SLO} The PDG value for $Br(J/\psi\rightarrow b_1^{\pm}\pi^{\mp}$) 
is $(3.0\pm0.5)\times 10^{-3}.$  As a check of our $\psi(2S)$ measurement,
we applied the same procedures to the $J/\psi\rt b_1\pi$ decays  
from $\psi(2S)\rt\pipi J/\psi$ cascade decays and find a
result that agrees within errors with the PDG value.
\bibitem{WMD} The function used to fit the $K_1(1270)$ is a
convolution of an S-wave Breit Wigner
function with mass and width fixed at 
the PDG values for the $K_1(1270)$
with a P-wave Breit Wigner function for the $\rho$ meson.   
The requirement
that the $\pi^+\pi^-$ mass is within $\pm150$~MeV of the $\rho(770)$
is set in the limits
of the convolution integral.
\bibitem{contamination} 
The $K^*\pi$ feedthroughs are taken into account 
both for $\psi' $ and
$J/\psi \rt K1(1270) \overline{K}$, and have
efficiencies respectively of $0.032 \pm 0.005$, 
and $0.010 \pm 0.002$. 
The errors on all efficiencies 
include the errors from MC statistics and systematics added in 
quadrature.
\bibitem{limit} The 90\% CL limits are obtained by increasing the
number of events from the fits by $1.28~\sigma$, where
$\sigma$ includes the error from the fit and the 
systematic error added in quadrature.
\bibitem{1400fdt} The spectrum is fit taking into account 
the $K_1(1400) \rt K^*\pi$ feedthrough into the
$K \rho $ channel.
\end{references}
\end{document}